\documentclass[lettersize,journal]{IEEEtran}
\usepackage{amsmath,amsfonts}
\usepackage{algorithmic}
\usepackage{algorithm}
\usepackage{array}
\usepackage{subcaption}
\usepackage[caption=false,font=normalsize,labelfont=sf,textfont=sf]{subfig}
\usepackage{textcomp}
\usepackage{stfloats}
\usepackage{url}
\usepackage{verbatim}
\usepackage{graphicx}
\usepackage{cite}
\usepackage{multirow}
\usepackage{import}

\usepackage{graphicx}

\begin{document}

\title{Exploiting HDMI and USB Ports for GPU Side-Channel Insights}

\author{\IEEEauthorblockN{Sayed Erfan Arefin}
\IEEEauthorblockA{\textit{Department of Computer Science} \\
\textit{Texas Tech University, Lubbock, Texas}\\
saarefin@ttu.edu}
\and

\IEEEauthorblockN{Abdul Serwadda}
\IEEEauthorblockA{\textit{Department of Computer Science} \\
\textit{Texas Tech University, Lubbock, Texas}\\
abdul.serwadda@ttu.edu}
}




\maketitle

\begin{abstract}

 Modern computers rely on USB and HDMI ports for connecting external peripherals and display devices. Despite their built-in security measures, these ports remain susceptible to passive power-based side-channel attacks. This paper presents a new class of attacks that exploit power consumption patterns at these ports to infer GPU activities. We develop a custom device that plugs into these ports and demonstrate that its high-resolution power measurements can drive successful inferences about GPU processes, such as neural network computations and video rendering. The ubiquitous presence of USB and HDMI ports allows for discreet placement of the device, and its non-interference with data channels ensures that no security alerts are triggered. Our findings underscore the need to reevaluate and strengthen the current generation of HDMI and USB port security defenses. 

\end{abstract}


\section{Introduction}
Today's computers are equipped with numerous Universal Serial Bus (USB) and High-Definition Multimedia Interface (HDMI) ports to support the wide range of external peripherals and visualization needs of modern users. These ports provide a physical channel to the underlying hardware and software, making them integral to the computer's threat surface. As such, they are built with security features designed to prevent exploitation. For instance, USB ports disable the autorun feature to prevent malware on a USB-connected device from executing automatically. Operating systems also notify users whenever a USB-connected device attempts to access the file system, allowing users to block unexpected actions \cite{autorun_literature}. HDMI ports, primarily output devices, do not support writing to the computer, further reducing their susceptibility to certain types of attacks. These and other security measures help safeguard these physical ports from becoming vectors for attacks \cite{hardware_details_1} \cite{hardware_details_2}.

In this paper we argue that current port security defenses are inadequate in the wake of passive attacks that leverage these ports as a vehicle to drive power measurements designed to infer key underlying processes on the computer. Using the example of processes running on the GPU, we demonstrate the viability of these attacks through the development of a custom device that plugs into USB and HDMI ports, facilitating inference attacks on GPU activities.

Our approach is stealthy and does not interfere with the processes that typically trigger user alerts. For the USB port, our attack is designed to avoid manipulating the data channel, thereby bypassing security notifications. For the HDMI port, our device mimics a conventional display device, ensuring it does not trigger any errors or alerts. The ubiquitous presence of USB and HDMI ports on contemporary computers -- located at the back, front, and sides -- facilitates the discreet deployment of our attack devices. A small, inconspicuous device plugged into any of these ports can remain unnoticed, providing an effective means of conducting attacks without installing software on the victim’s computer or attaching hardware to internal components.

The overarching idea behind our attack is as follows: despite modern computers being equipped with over-provisioned power supplies and sophisticated stabilization mechanisms to prevent power instability, significant power draw increases by a GPU during intensive processes can induce detectable fluctuations in power seen by peripheral devices attached to USB or HDMI ports. These fluctuations, though transient and eventually stabilized, contain discernible information sufficient to infer GPU operations. Using GPU loads in the form of matrix multiplications, video scene rendering and neural network execution, we demonstrate that this line of attack poses a potent side-channel that can leak sensitive information about processes running on the GPU.

\begin{figure*}[ht]
     \centering
         \includegraphics[width=\textwidth]{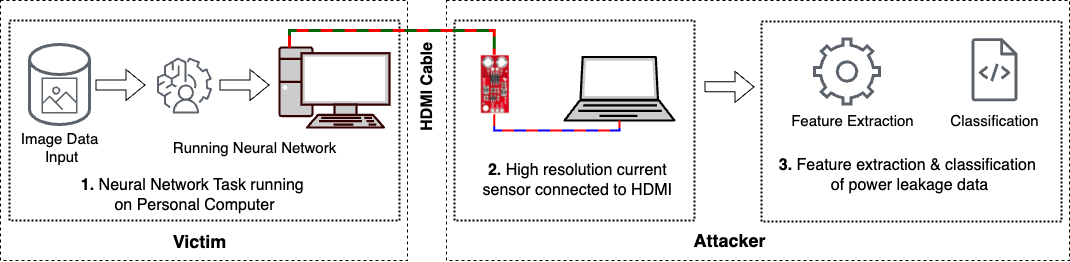}
        \caption{Illustration of the attack process. The example used here is that of a neural network running on the victim/targeted computer}
        \label{experiment-full-overview}
\end{figure*}
\subsection{Paper Contributions}
Our research makes the following Contributions:

(1) GPU power side-channel inference via HDMI and USB ports: Our research introduces a novel family of power side-channel attacks, which utilize power consumption patterns observed at USB and HDMI ports to infer computational processes occurring on a GPU. This method is distinct from traditional attacks that measure power directly from the GPU, as it exploits peripheral connections that either interface with or share a power source with the GPU. Requiring only the insertion of a device into a peripheral port, this stealthy approach operates at the user's privilege level without triggering any alerts. This technique not only broadens the scope of implementable side-channel attacks but also suggests the potential for using peripheral power measurements defensively, such as detecting unusual power draw signatures that indicate malicious activities.

(2) Design of power measurement prototype: We present the design of the power measurement tool central to our attack strategy. At the core of this tool is a SparkFun ACS723 Current Sensor, which utilizes a Hall effect sensor to produce a voltage output correlating to the current through its measurement pins. This sensor is integrated with an Arduino board equipped with a 32-bit ARM Cortex M0+ processor, enabling data collection at a rate of up to 350,000 samples per second. This high-resolution data is crucial for driving the machine learning classifiers that discern and classify GPU activities based on power fluctuations.

(3) Evaluation of attack performance: Our evaluation of the attack's performance begins with an intuitive analysis of visual patterns in time-series data extracted from GPU computations. This preliminary analysis provides initial indications for the feasibility of the attack, leading to the development of machine learning models capable of executing two distinct attack scenarios with high accuracy. In the first scenario, our side-channel approach successfully identifies the specific family of deep neural networks running on the GPU. In the second, it accurately infers which video clips are being rendered by the GPU. These examples demonstrate the broader applicability and effectiveness of our side-channel techniques in targeting GPU-based processes.

\subsection{Threat Model}
\label{subsec:threat-model}
Our threat model assumes an attacker who has physical access to a target computer but lacks login credentials. Such individuals might include office cleaners, maintenance staff, or rogue employees who can access the machine without raising suspicion. The attacker can discretely insert a small power measurement device into an inconspicuous USB or HDMI port on the computer, preferably one that is not easily visible (e.g., the ports at the back). This device is designed to operate stealthily, drawing power data without causing any disruptions or triggering the alerts typically associated with peripheral connections like USB storage devices. The device does not interfere with normal computer operations, such as display functions when connected to an HDMI port. 

The attacker's goals may vary from stealing power measurement data linked to high-value algorithms, such as neural network architectures or cryptographic operations, to general surveillance of the user's GPU-dependent activities, which might include video rendering or website browsing. To facilitate the machine learning aspect of the attacks, the attacker would need to acquire training data that reflects the specific computational processes targeted.

In our laboratory setup, we utilized a wired connection for data transfer, sufficient for demonstrating the attack's foundational concepts. However, in a practical application, this setup could be easily modified to include wireless data transfer, such as using Bluetooth or Wifi to transmit data to a device in a nearby location. Figure \ref{experiment-full-overview} captures the attack process, taking the example where the attacker seeks to infer information about a neural network architecture running on the victim's device.

\subsection{HDMI/USB Power Side-Channel for Defense}
While the primary focus of our paper is on exploiting the described side-channel for attacks, it is important to note its potential defensive applications. Specifically, this technique could be useful for anomaly detection or malware identification through deviations in power consumption patterns. For example, a computer infected with malware that engages in resource-intensive activities unknown to the user --- such as crypto-mining, data exfiltration, or unauthorized network attacks --- would exhibit abnormal power usage. Employing a USB-type or HDMI-type power measurement device can provide a significant advantage in such scenarios. It operates independently of the computer's software, making it less susceptible to tampering by malware that might disable or manipulate software-based monitoring tools. Furthermore, its simple installation and ease of use make it accessible to users without specialized technical skills, offering a non-invasive, reliable method for monitoring system integrity.




\section{Related Works}
\label{sec:related-works}
Our research has two clusters of related work: power side-channel attacks on computing devices, and attacks that exploit vulnerabilities of I/O devices. In this section, we discuss these two families of research and describe how we differ from them. 

\subsection{Power side-channel attacks:} The wide array of past studies on power side-channel attacks measured power in several different ways depending on the threat scenario and targets of the attack. We categorize these related works based on the ways in which power was measured before discussing how our work differs from this body of past research. 

\textbf{Using Nvidia-smi interface:} 
Nvidia-smi is a widely used interface to measure nvidia GPU resources, including power consumption \cite{nvsmi}. The interface utilizes the nvml library \cite{nvml} provided by Nvidia GPUs. The research conducted by Jha et al. \cite{10.1145/3414552} investigates a two-stage attack methodology designed to infer the architecture of deep neural networks (DNN). The study utilizes nvidia-smi to analyze performance on P100 and P4000 GPUs, demonstrating the attack's effectiveness. The authors proposed a secure MobileNet-V1 as a mitigation to such vulnerability. In their work Sabbagh et al. \cite{9218690} presents an overdrive fault attack on modern GPUs, exploiting nvidia-smi to introduce random faults during kernel execution. The authors effectively recovered AES keys in their experiments. The study by Sayed et al. \cite{aiiot} explores power consumption patterns using nvidia-smi to determine if this data could reveal key details about DNN architectures running on GPUs. They conducted model inference attack on several nvidia GPUs including some cloud GPUs.

\textbf{Using Intel's Running Average Power Limit (RAPL) interface:} RAPL is the primary interface for power measurement on Intel CPUs. 
Moritz et al. \cite{9519416} introduced a software-based power side-channel attack that leverages unprivileged access to this interface. The authors successfully extracted cryptographic keys and monitored application control flows without needing physical access or specialized hardware. Demonstrations included breaching Intel SGX to extract AES-NI keys, and recovering RSA keys via privileged attacks. Xiang et al., \cite{10323806} conducted a research where they introduced a novel approach to model extraction in deep learning using the Intel Running Average Power Limit (RAPL) to exploit power leakage in ReLU activation functions via a software interface. Their attack used a very small number of queries to extract deep learning models by leveraging software-based power side-channel information. This technique was implemented on the oneDNN framework. They extracted a 5-layer MLP and Lenet-5 CNN. The study conducted by Segev et al., \cite{10.1007/978-3-031-34671-2_16} showed how performance counters can be used to infer information about neural networks. They used the Intel Power Gadget to collect detailed performance traces, including power consumption and CPU utilization. The software used the Intel RAPL technology in its core.

\textbf{Exploiting Dynamic Voltage and Frequency Scaling (DVFS):}
DVFS is a power management mechanism used in Intel, AMD, and ARM CPUs, that dynamically adjusts CPU frequency and voltage to reduce overall system power consumption and enhance performance per Watt. While DVFS helps in managing energy efficiency, it also introduces a potential vulnerability that can be exploited through power side-channel attacks.
For example, by observing the power consumption variations that occur as a result of DVFS adjustments during the computation process. Chen et al., \cite{Liu_2022} successfully extracted AES keys by leveraging the DVFS mechanism, while Yingchen et al., \cite{10122602} studied how power side-channel attacks on modern Intel and AMD x86 CPUs can be used via Dynamic Voltage and Frequency Scaling (DVFS), without requiring access to a power measurement interface directly. Debopriya et al., \cite{10.1145/3564625.3567979} used the DVFS side-channel attack to perform website fingerprinting for browsers like Google Chrome and Tor across various CPU architectures, while Sehatbakhsh \cite{9065580} manipulated the power management units to enable side-channel attacks through electromagnetic emanations from the voltage regulator module (VRM). They utilized this technique to execute keystroke logging.

\textbf{Direct power measurement using an oscilloscope:}
Hasindu et al., \cite{gamaarachchi2018power} outlines a power analysis attack, where they focus on extracting cryptographic keys from devices by analyzing power consumption data during cryptographic operations. The researchers demonstrate direct power measurement using an oscilloscope. They utilize power analysis techniques to break the Speck algorithm for embedded systems. Luo et al., \cite{7357115} study the power side-channel on Graphics Processing Units (GPUs) to extract a secret key from a block cipher operating on a GPU. In order to capture power they inserted a resistor in series with the PCI-Express power supply and measured the power physically. They targeted a CUDA AES implementation on an Nvidia Tesla GPU.

\textbf{Phone charging stations:}
Yang et al., \cite{7782756} highlighted the privacy risks posed by public USB smartphone charging stations whose public placement makes them vulnerable to malicious entities that rigg them with power measurement functionality. Using one such rigged hub, the researchers demonstrated a side-channel attack that can identify web pages loaded on a smartphone during the phone charging process. They conducted their research with different browser settings, network types, and battery levels, revealing successful attacks under all these settings. Sraddhanjali et al., \cite{9604475} extended this line of attack and demonstrated that USB charging hubs can analyze power consumption data to infer YouTube music videos that a user is watching on their smartphone, while Alexander et al., \cite{10.1145/3460120.3484733} focused on the wireless charging interfaces on modern smartphones. They showed that the wireless charging interface is also susceptible to power-side-channel attacks, specifically demonstrating a website fingerprinting attack on iOS and Android devices. 

\textbf{Other power measurement approaches:}
Zhao et al., \cite{8418606} showcased a new security vulnerability introduced by integrating Field Programmable Gate Arrays (FPGAs) into cloud data centers and System-on-Chips (SoCs). Their work reveals that FPGAs can facilitate software-based power side-channel attacks remotely without the need for physical access to the target system. The researchers demonstrate using ring oscillators on FPGAs to monitor power consumption of both FPGA modules and CPU components within the same SoC, successfully executing power analysis attacks on RSA cryptomodules and bypassing timing-channel protections for CPUs. Sayed et al. \cite{10.1145/3437880.3460415} demonstrated a power side-channel attack on the GPU to infer deep neural network architectures using a software-based approach. Their attack reads power consumption data from the on-board sensors of the GPU.

{\bf{How we differ from these works:}}
The fundamental difference between these studies and our research is that we introduce a novel side-channel -- specifically, power leakage through peripheral devices such as the HDMI and USB ports. While previous works have relied on either invasive physical measurements directly from the power supply or sophisticated software approaches that require specific system access or vulnerabilities, our approach works by methodically triggering a power draw from the HDMI/USB ports that then enables inferences on the underlying processes running on the GPU. This is achieved without altering the device or needing special privileges. 

Notably, our method capitalizes on the inherent power variations that pass through these peripheral ports as a byproduct of GPU processing loads, making it a unique point of observation that remains largely under-explored in current literature. Our technique, therefore, opens new avenues for both understanding and mitigating security risks in modern computing environments where external peripherals are common.

On the surface, the charging station attacks (e.g., \cite{7782756, 9604475, 10.1145/3460120.3484733} might seem to have some similarity to our work given that they also use a USB port. However, that attack is fundamentally different from our attack since the target of the attack  is the phone plugged into the USB port, and the power being drawn by the phone is directly measured by a circuit placed between the phone and the USB port. Inferences about activities being undertaken on the phone are then done based on these direct power measurements. In our attacks on the other hand, the USB (and HDMI) ports are simply a subset of the many peripherals that share a power source with the GPU, indirectly picking up GPU power patterns through fluctuations seen in its own power as supplied by the source. These indirect power patterns are then used to infer information about computations that occur on the GPU.

\subsection{I/O port related vulnerabilities:}
The body of past research on I/O security vulnerabilities related to our research is mainly focused on the USB port, given its usage for a wide range of device connectivity applications. We only briefly review these works, given that they do not focus on the power side-channel which drives our work. 

Ramadhanty et al. \cite{9274631} investigate the vulnerability of Windows 10 to USB-based keylogger attacks. By embedding a PowerShell script in an Arduino Micro, they demonstrated a successful keyboard injection attack. Pham et al. \cite{PHAM2011172} examine the information security risks introduced by USB storage devices due to their insecure design and open standards. Their work reviews various USB-based software attacks on host computers and USB devices, exploiting vulnerabilities in USB protocols, embedded security software, drivers, and Windows Autoplay features. 

Nohl et al. \cite{Nohl2014} investigate an attack in which USB firmware is reprogrammed to bypass traditional security measures and execute unauthorized commands, while Jeong et al. \cite{4547620} analyzes the vulnerabilities of popular secure USB flash drives. They found that passwords can be exposed during communication between the USB drive and the PC. Furthermore, the study reveals that some secure USB flash drives are susceptible to firmware attacks, allowing unauthorized users to bypass security measures. 

Brian et al. \cite{anderson2010seven} show how the USB Switchblade could be used to steal sensitive information such as passwords, network configurations, and system details, making it a powerful tool for attackers to gain unauthorized access and steal data from a target machine. Kaspersky Lab \cite{greatequation} discusses scenarios in which a compromised USB storage device can be used to secretly extract sensitive data from a victim machine. This process occurs without the awareness of the host or the USB device owner. 

Tariq et al. \cite{Tariq2023} review cybersecurity challenges in IoT, focusing on vulnerabilities in USB-connected devices. They discuss the risks of malware spreading through USB ports and emphasize the need for securing these interfaces. The authors highlight the importance of developing robust security measures to mitigate these threats, suggesting the use of better authentication protocols and noting that traditional security measures are inadequate for modern IoT environments.

{\bf{How we differ from these works:}} While previous works share our objective of highlighting the security threats posed by USB devices, our approach is distinct in that it focuses on a passive power measurement attack. Unlike these prior attacks, our method does not require any special privileges or engage in any communication with the underlying operating system or applications. Instead, it passively measures power consumption patterns, making it a stealthier and less intrusive method compared to attacks that actively probe the target system.

\section{Basics of Computer Power Supply Mechanisms}
\label{sec:preliminary}
To provide context to our attack design, this section covers some basic information about the power supply mechanism is computers. We first discuss the power supply unit (PSU) and how power is supplied to different peripherals of the computer. We finally discuss some input and output ports (I/O ports) that can potentially power external devices.

\begin{figure}[!ht]
    \centering
    \begin{subfigure}[b]{0.45\textwidth}
        \centering
        \includegraphics[width=\textwidth]{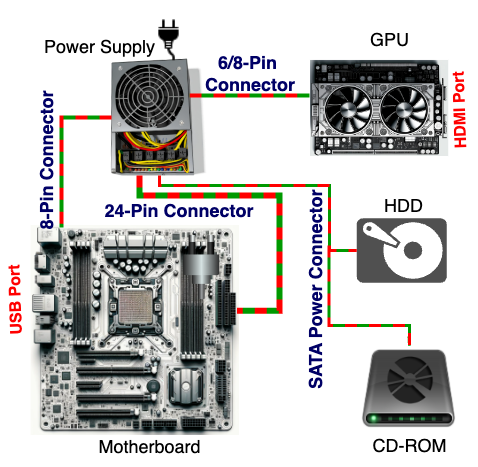}
        \caption{How power is supplied inside the computer components.}
        \label{power-supply-image-2}
    \end{subfigure}
    \hfill 
    \begin{subfigure}[b]{0.5\textwidth}
        \centering
        \includegraphics[width=0.4\textwidth]{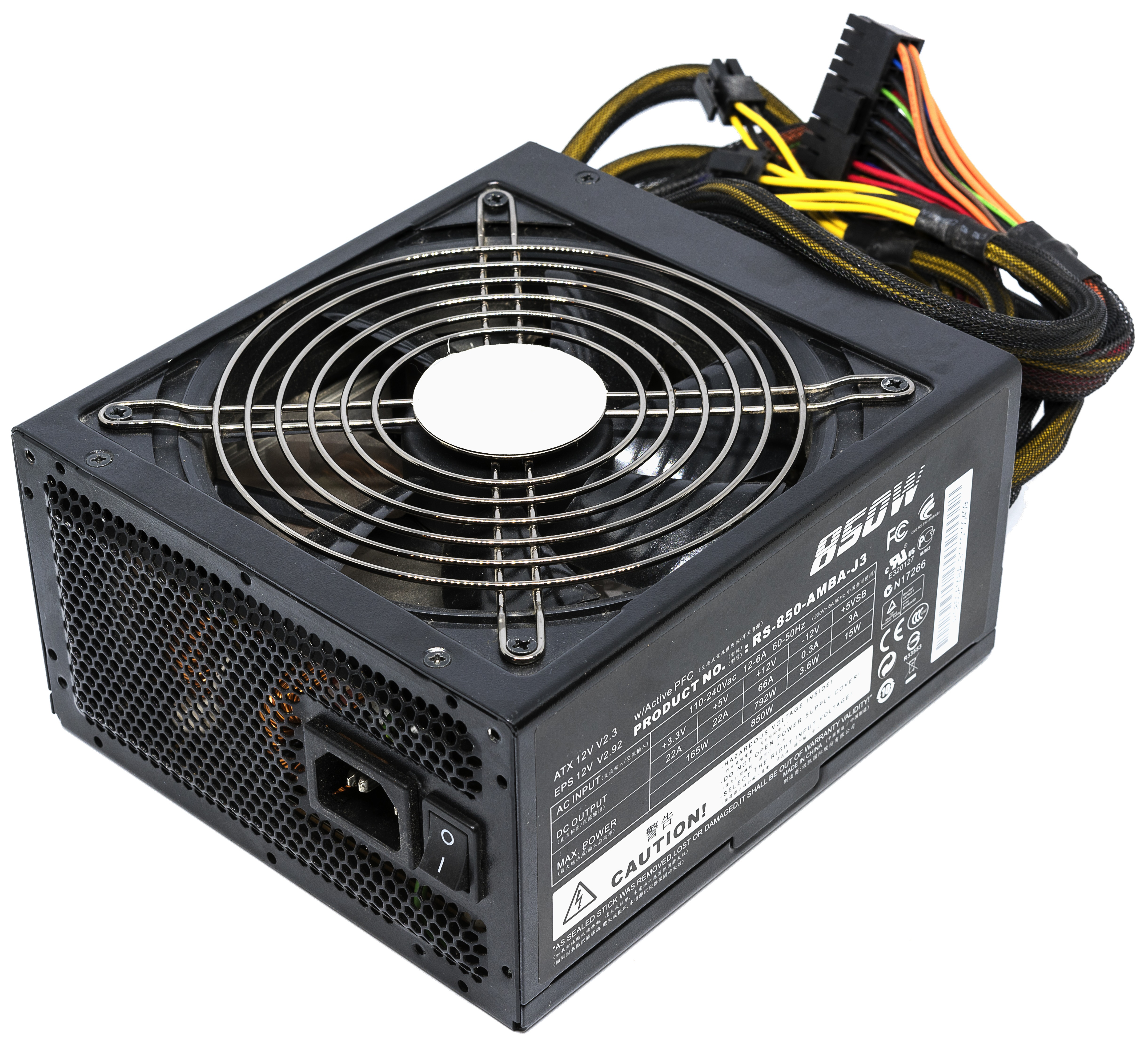}
        \caption{A sample ATX Power Supply}
        \label{power-supply-image}
    \end{subfigure}
    \caption{Desktop Computer Power supply.}
    
    \label{power-supply-related}
\end{figure}

\subsection{Computer Power Supply Unit (PSU)}
Often the power supply used in a home-usage, gaming computer or a workstation is an ATX power supply. A sample picture of an ATX power supply is shown in Figure \ref{power-supply-image} \cite{satrinekarn2023power}. Various power cables are used to connect all the components to the power supply. Each type of connector has specific current and voltage ratings. A very widely used connector is the 24-pin ATX power connector, which connects to the motherboard. It can deliver multiple voltages, such as +3.3V, +5V, and +12V. The 24-pin connector can supply significant current, up to 24A on the +5V rail and 24A on the 3.3V rail. However, this current supply depends on the power supply's capacity. The SATA power connector is used to power devices with a SATA interface such as hard disk drives (HDDs), solid state drives (SSDs), or optical drives (CD/DVD drives). This connector supplies three voltages: 3.3V, 5V, and 12V. The CPU power connectors, typically known as the 8-pin connector, connect to the motherboard to supply power to the main processor. This connector can provide +12V to the CPU. Today, most computers are equipped with high-performance graphics cards. These are used for advanced graphics processing and machine learning tasks. The 6-pin or 8-pin PCIe connector can deliver high power to GPUs \cite{atx_power_supplies}. An illustration of the power supply connectors and the corresponding computer components is presented in Figure \ref{power-supply-related}.


\subsection{Computer I/O ports}
There are several input and output (I / O) ports in the computer that can be used to connect different peripherals. These peripherals can be a wide variety of devices. In case of USB devices, keyboard, mouse, external hard disk drive, etc. are a few notable peripherals. These devices require power to operate, which is consumed over the USB port. The current flow from a regular USB port can vary depending on the specific USB standard and the device providing the power. For USB 2.0, the standard current output is typically 500 mA (milliamperes) at 5 volts, providing a maximum power of 2.5 watts. For USB 3.0 and later versions, the current output can be higher, often up to 900 mA or more, again at 5 volts. In our experiments, we used the USB 2.0 ports. 

We also observed the current draw made by the HDMI port. The HDMI port is in several ways a more complicated port than the USB port. It does not supply active power to the display devices, as those require external power. However, it has a power saving option, where the device goes to sleep if there is no signal in the HDMI port. This is called hot plug detection (HPD). There is a small circuit that periodically checks for the signal such that, when there is a signal, the display device can come up from its sleep status. In order to run this small circuit, a small amount of power is supplied over the HDMI port. Pin 18 of the port supplies 5v and at most 500 milliampere, to run this circuit. In our experiments described in Section \ref{sec:attack-experimets} we monitor the 18th pin of the HDMI port and the power pin of the USB port. Details of the sensor and circuit used to make measurements on this pin are discussed in Section \ref{sec:sensor_and_curcuit}.

\section{Attack Design}
\label{sec:attack-design}

\begin{figure*}[!htb]
     \centering
         \includegraphics[width=0.7\textwidth]{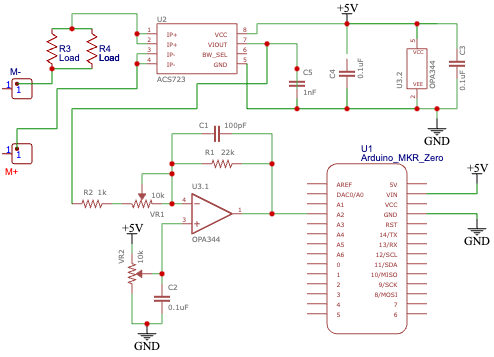}
        \caption{Circuit of the attack device.}
        \label{circuit}
\end{figure*}

In this section of the paper, we discuss our hardware setup to measure current and possible power leakage from the I/O ports. The following sections describe the circuits used and the hardware setup in Section \ref{sec:sensor_and_curcuit}. High resolution data collection required certain setup for the analog-to-digital converter (ADC), which is described in Section \ref{sec:arduino-implementation}. In addition, the computer used to run our experiments is described in Section \ref{sec:experiment-computer}.

\subsection{Hardware setup}
\label{sec:sensor_and_curcuit}

At the heart of our hardware lies a current sensor capable of recording the current passing through a series circuit. The hardware setup incorporated a SparkFun ACS723 Current Sensor \cite{sparkfun14544}. This module uses a Hall effect sensor to generate a voltage output that corresponds to the current passing through the measurement pins on the board. Since it uses a hall effect sensor, it ensures electrical isolation between the monitored circuit and the measuring circuit. The module can measure a current from 10mA to 5A. It has an onboard preamp to adjust the resolution of the readings. We had increased this to an optimum point to have a better resolution of the reading. The analog output of the module is set to 20kHz in order to reduce noise as the gains are set to high. The sensor is connected to an Arduino to read the analog signal and produce a digital reading to be used later with machine learning classifiers. In our experiments, we used an Arduino MKR Zero board. This board is equipped with an Atmel SAMD21, which uses a 32-bit ARM Cortex M0+ processor\cite{mkr_zero}. 

The analog-to-digital converter (ADC) can achieve sampling rates as high as 350,000 samples per second. The sensor has 3 pins. These are the +5V, ground, and analog output pins. The analog output pin of the sensor is connected to the analog input pin (A2) of the arduino. The ground and 5V are connected to the corresponding power terminal of the circuit. The sensor board is equipped with an opamp and other operational components. The final assembly of the whole circuit can be seen in Figure \ref{circuit}. In order to measure the current, we need to place the current sensor in series. But, since we want to measure the current draw of the I/O ports, we cannot put it in series directly, but rather have a load in the series. Without the load, the short-circuit protection of the I/O boards may trigger. 

\begin{figure}[!htb]
     \centering
         \includegraphics[width=0.18\textwidth]{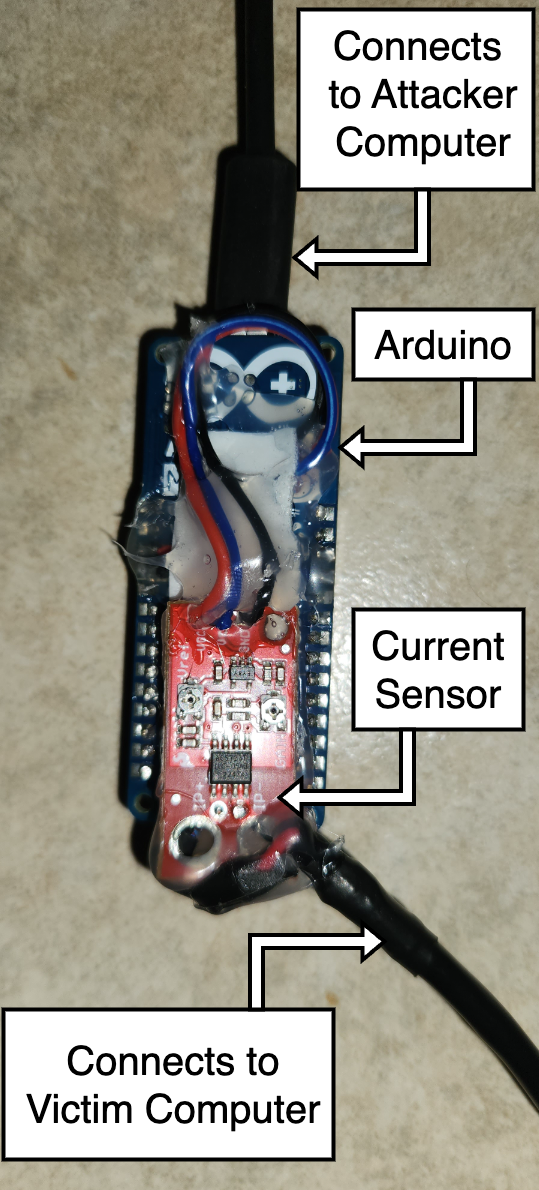}
        \caption{The attack device.}
        \label{the-device}
\end{figure}

Power dissipation in resistors generates heat. Distributing the power dissipation across multiple resistors can help to better manage heat, preventing any single resistor from overheating. In our experiment, we had two resistors in parallel. Each resistor in our experiment was a $20\Omega$ film resistor. These resistors are marked as load R3 and R4 in the circuit presented in Figure \ref{circuit}. The final assembly of our attack device is presented in figure \ref{the-device} which labels the components of the device. The placement of the current sensor on top of the Arduino can be observed. The figure also labels two cables. One will be connected to the victim computer to measure power. The other cable is connected to the attacker computer. 

In Section \ref{subsec:threat-model}, we discussed the possibility of enhancing the attack device by integrating a Bluetooth or WiFi module, thereby extending the operational range of the attacker. In this advanced scenario, the cumbersome cable to the attacker device becomes obsolete, replaced by seamless wireless communication. The attack device, in this extended setup, would draw power from an independent source, such as a 5V battery.


\subsection{Arduino Implementation Details}
\label{sec:arduino-implementation}
In our Arduino software setup, we perform high-speed data acquisition from an analog current sensor. Our objective is to capture the analog data and transmit them to a computer, where they are saved as text files (.txt) using a Python script. To facilitate rapid data transfer, we configure the Arduino's serial communication to operate at a baud rate of 2,000,000 bits per second. We adjust the Analog-to-Digital Converter (ADC) to optimize data quality and sampling rate. The ADC resolution is set to 12 bits to enhance the fidelity of the digital conversion. As mentioned earlier, the ADC samples from the analog pin A2, which connects to the current sensor's output. Our code uses continuous sampling with interrupts, allowing the ADC to function efficiently with minimal processor overhead. We configure the ADC’s control register B (CTRLB) to set the clock prescaler to a 40 kHz operating frequency. This configuration achieves a sampling rate of 40,000 samples per second, which is well above the required 20,000 samples per second for our current sensor application. An ADC Interrupt Service Routine (ISR) is programmed to handle the data conversion results. It reads the ADC output, transmits it via USB, and resets the ADC for the next conversion.

\subsection{Experiment Machine}
\label{sec:experiment-computer}
The experiments were carried out on a desktop computer that had the following configuration. The graphics processing unit (GPU) used in the computer is an Nvidia RTX 2060 Super with 8GB graphics memory. This uses the Nvidia Turing Architecture. The other components of the computer include a Central Processing Unit (CPU) Intel Core i7 9600K Processor, 16 GB RAM, 4 TB hard drive (HDD). The operating system (OS) used is Ubuntu 22.04 LTS. The tensorflow version used in the experiment is a tensorflow-gpu 2.12. The computability of the tensorflow setup is 7.5. The power supply used in the computer is a 450 watt power supply. The GPU uses a 6-pin power connector. GPU requires a 175 Watt power and CPU requires a 95 Watt power. The other components draw a variable power based on the need.

\section{Attack Experiments}
\label{sec:attack-experimets}
In this section of the study we discuss different parts of the attack experiments. At first we conduct several preliminary experiments and observe the power leakage phenomenon (Section \ref{seubsec:preliminary-experiments}). Afterwards, we discuss how we conducted data collection and experiments to infer the Convolutional Neural Networks model architecture with several levels of experiment configuration (Section \ref{subsubsec:infer-cnn-configs}). We also performed experiments in inferring video from several video-rendering tasks (Section \ref{subsubsec:infer-video-configs}). In the later part of this section, we discuss the datasets used in our CNN model architecture inference experiment (\ref{sec:cnn-datasets}) and video inference experiment (\ref{sec:video-datasets}).

\subsection{Preliminary Experiments}
\label{seubsec:preliminary-experiments}

Before undertaking a detailed design of the attack, we first conducted a preliminary experiment to discern if GPU operations produce measurable current draw fluctuations at our attack device. In the first of these preliminary attacks, we used the Glmark 2 software \cite{glmark2} to render videos on the GPU. This software spawns a window and renders scenes that make use of OpenGL (ES) 2.0 and is commonly used to benchmark GPU performance. Before collecting any data, we ran the command to see GPU activity using the Nvtop tool \cite{nvtop} and observed some GPU utilization. Figure  \ref{fig:features_mean_mm} captures the findings from this experiment for one of the scenes. The figure reveals a slight hump in between the 1 and 2-second marks when the rendering happens. For both the USB and HDMI ports, our device is hence able to pick up the power fluctuations triggered by the video rendering on the GPU.  Similar patterns (not shown here) were observed for various other scenes, providing us with the first signs that our line of attack might be feasible.



\begin{figure*}[!h]
    \centering
    \begin{subfigure}[b]{0.94\textwidth}
        \centering
        \includegraphics[width=1\linewidth]{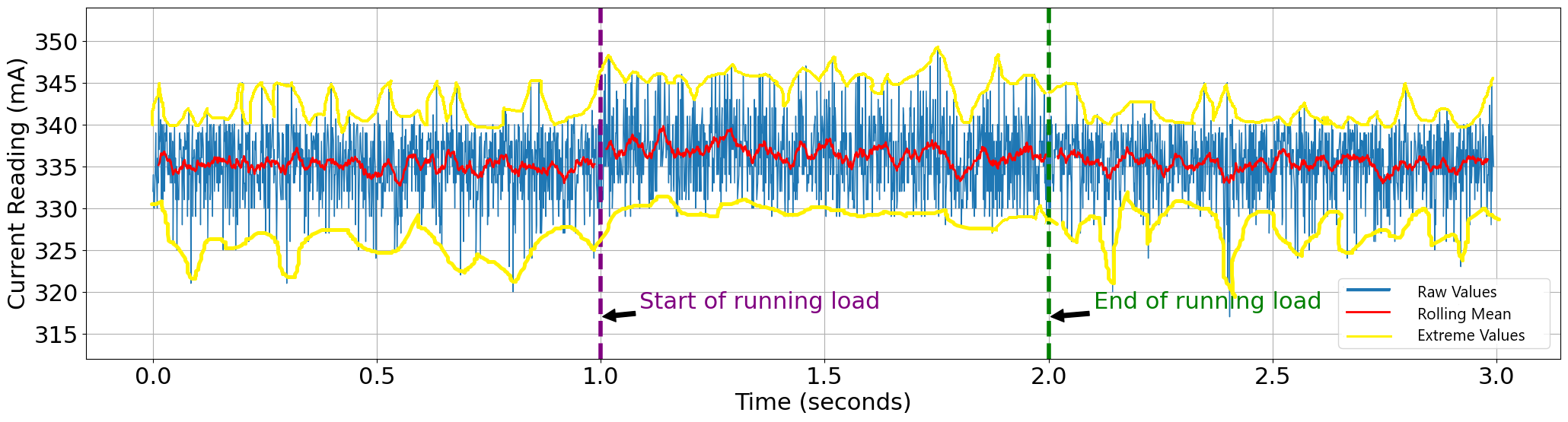}
    \caption{HDMI}
   
    \end{subfigure}
    \vfill
    \begin{subfigure}[b]{0.94\textwidth}
        \centering
        \includegraphics[width=1\linewidth]{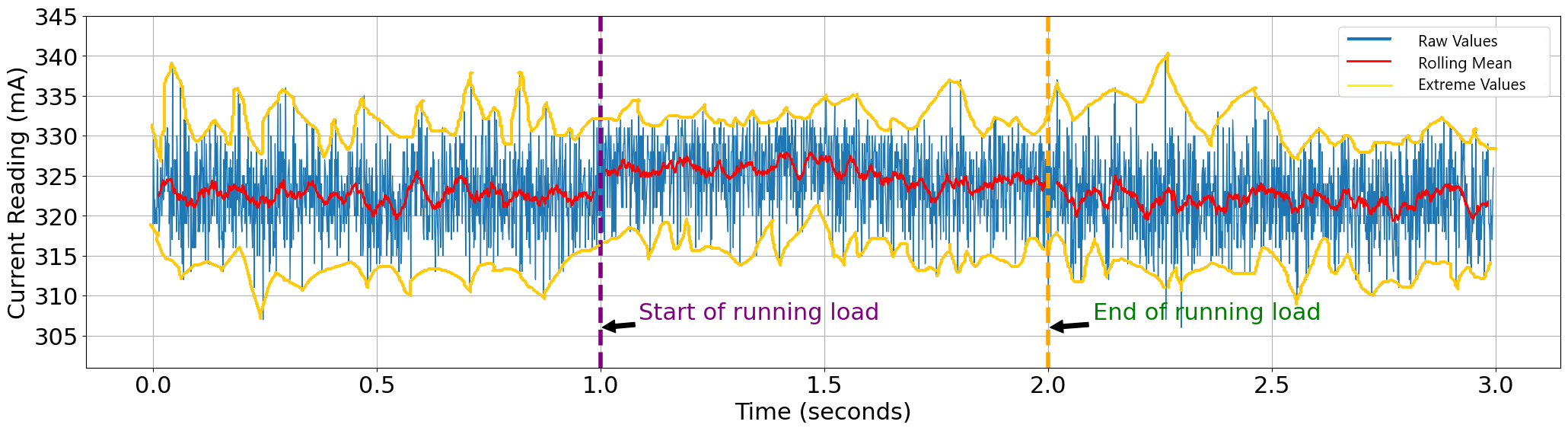}
        \caption{USB}
    \end{subfigure}
    \caption{Raw reading for Scene rendering with Glmark software.}
    \label{fig:features_mean_mm}
\end{figure*}

Our second preliminary experiment focused on matrix multiplication, a fundamental computation underlying many GPU operations such as image processing and neural network computations. We aimed to observe power leakage from the I/O ports during different types of matrix multiplication.

To formulate this experiment, we used matrices of size 1920 x 1080, matching the resolution of a typical full HD display. This setup mimics a regular use case where a matrix of this size is executed on the GPU. The goal was to multiply two matrices on the GPU, ensuring execution via the CuPy library \cite{cupy}, which utilizes the GPU for matrix manipulation operations.

We created four matrices, A, B, C, and D, with random values between 0 and 255 to replicate the color values of each pixel. We then performed two matrix multiplication tasks: A multiplied by B, and C multiplied by D. Each multiplication task was repeated 25 times to observe the power leakage phenomenon. Following the multiplications, we extracted features, specifically the mean and MFCC 2 (Mel-frequency cepstrum coefficient) of the time series data, and plotted these features.

This experiment was conducted for both HDMI and USB I/O ports. The feature plots are presented in Figure \ref{fig:features_mean_mm}. The plots show that our USB and HDMI measurements depict distinct clusters of power measurement data for the two different matrix multiplication tasks conducted on the GPU. Given these positive results from the two toy experiments, we proceeded to design and execute the fully-fledged attack experiment that we discuss next.

\begin{table}[h!]
\centering
   \resizebox{0.495\textwidth}{!}{
\begin{tabular}{|c|c|c|}
\hline
 \textbf{Base Name} & \textbf{Variants} \\
\hline  \hline  
 ConvNeXt & ConvNeXtBase, ConvNeXtLarge, ConvNeXtSmall \\
\hline
 VGG Net & VGG16, VGG19 \\
\hline
ResNet & ResNet50, ResNet152V2, ResNetRS420 \\
\hline
InceptionNet & InceptionV3, InceptionResNetV2 \\
\hline
 MobileNet & MobileNet, MobileNetV2 \\
\hline
DenseNet & DenseNet121, DenseNet169, DenseNet201 \\
\hline
 NASNet & NASNetMobile, NASNetLarge \\
\hline
RegNet & RegNetX002, RegNetX160, RegNetY320 \\
\hline
EfficientNet & EfficientNetB0, EfficientNetB5, EfficientNetV2L \\
\hline
XceptionNet & (No variants) \\
\hline
\end{tabular}
}

\caption{Base CNNs and their Variants used in our Experiments}
\label{table:network_models}
\end{table}

\begin{figure}[!ht]
    \centering
    \begin{subfigure}[b]{0.235\textwidth}
        \centering
        \includegraphics[width=\textwidth]{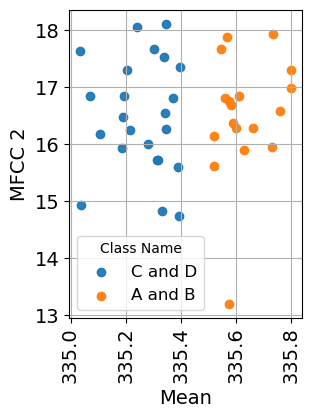}
        \caption{HDMI}
    \end{subfigure}
    \hfill
    \begin{subfigure}[b]{0.235\textwidth}
        \centering
        \includegraphics[width=\textwidth]{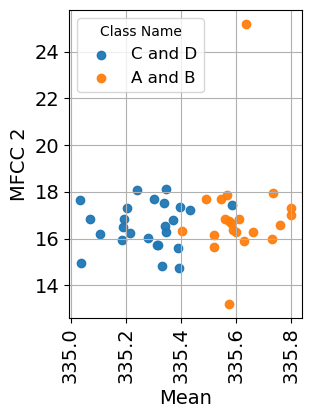}
        \caption{USB}
    \end{subfigure}
    \caption{Plots of Mean feature and MFCC 2 for Matrix Multiplication experiments.}
    \label{fig:features_mean_mm}
\end{figure}

\subsection{Attack Configurations and Data collection}
In this part of the research, we focus on the detailed experiments of our research. We first run experiments to infer Convolutional Neural Networks and later infer video rendering.

\subsubsection{Inference of CNN Model Architecture running on the GPU}
\label{subsubsec:infer-cnn-configs}
Here we focus on the inference of the architecture of the convolutional neural network (CNN) using the power monitoring from the I/O ports. The CNNs are now widely used in many applications for a wide range of applications such as Image classification, Object detection, Video analysis, etc. The victim runs a CNN to classify the image, and the attacker will monitor the I/O port and try to infer the architecture of the CNN. We conducted this experiment for both HDMI and USB I/O ports. In this meticulously designed experiment, we postulate that the attacker has ingeniously adapted well-established CNN models and has proficiently trained machine learning models using the I/O port power data. Upon acquiring I/O port power data from the victim's machine, the attacker is equipped to precisely classify the specific CNN that was operational on the victim's system. Our investigation focuses primarily on the CNN architectures listed in Table \ref{table:network_models}, all of which benefit from the availability of pre-trained models in Tensorflow \cite{tf_modules_pretrained}. 

The experimental procedure begins with vigilant I/O port monitoring, followed by the beginning of CNN classification using the image data sets delineated in \ref{sec:datasets}. To ensure meticulous data integrity, a deliberate interval of 10 seconds is observed between each data collection sample. Recognizing the potential utilization of alternative CNN models by the victim, our experimental framework is also equipped to detect the presence of any unidentified model classes. We meticulously crafted our experimental design into two distinct configurations, Configuration 1 and Configuration 2, to probe the depths of CNN architecture inference. In Configuration 1, the attacker is armed with prior knowledge of the target CNN model's power consumption patterns, which are included in his training dataset. This setup allows for a direct and informed inference of the CNN model. Moving to Configuration 2, we introduced a layer of complexity where the attacker is only aware of the power data for one variant of the CNN model. This strategic limitation challenges the attacker to deduce the specific family to which an unknown variant of the CNN model belongs, based on partial data. This approach enhances our understanding of how well an attacker can generalize from known models to detect novel yet related model architectures. The data sets used in this experiment, which serve as input to CNN, are thoroughly described in Section \ref{sec:cnn-datasets}.

\subsubsection{Inference of Videos Rendered on the GPU}
\label{subsubsec:infer-video-configs}
Now a days, creating videos for the purpose of entertainment, video blogging (vlogging), or education is a very common task. One of the major use cases of the GPU of a computer is rendering videos. We further conduct an experiment evolving around video rendering. We design this experiment where the attacker wants to identify some video which is being rendered on the victim's machine. To run the experiment programmatically, we utilized the OpenCV library \cite{opencv_library} with Python. Here we use 10 different videos from different sources and try to classify the video. We assume that the attacker can train machine learning models with power leakage data for different videos. If such a video is rendered on the victim machine, the attacker can identify that by observing the power leakage data over the HDMI port of the victim machine. In this experiment, each video is cut for 10 seconds after the video is played for 10 seconds. We repeat this process for all the videos and 25 times each. 

\subsection{Experiment Datasets}
\label{sec:datasets}
In the following section, we present a detailed overview of the datasets utilized in our experiments. We describe the characteristics of these datasets, emphasizing their relevance for our experimental objectives. 

\subsubsection{Image Datasets for testing CNNs}
\label{sec:cnn-datasets}
The Convolutional Neural Networks (CNNs) typically take an image as input to perform different tasks (such as classification). This section discusses the image datasets used in our experiments for the CNNs. The images were taken from the ImageNet dataset \cite{5206848}. The ImageNet dataset is widely used in the ImageNet competitions, which has produced several CNNs over the years, such as VGG, ResNet, etc. We used 2 different categories of images. Dataset 1 contains the images of 50 cats and 50 dogs. Cats and dogs are typically misclassified among themselves. Dataset-2 contains 100 random images. We have further divided the data sets into two parts. In the later part of our experiment, we classify the CNN architectures -- inferred from the collected Power leakage. We had 80 images selected for the train set and 20 images for the test set. There were no common images between the train and the test sets.

\subsubsection{Video Inference Experiment}
\label{sec:video-datasets}

In this experiment we need to use video files as input to video rendering experiment. In the process of determining the appropriate format and quality of the video utilized in the Video Identification Experiment, the ubiquity of smartphones as recording devices was considered. Contemporary smartphones generally possess the capability to record in Full HD (1080p) or superior resolutions. The mp4 file format, commonly employed for the storage of video content, coupled with the H.264 video codec, strikes an optimal balance between superior video quality and minimized bandwidth consumption, rendering it exemplary for streaming and broadcasting applications. Consequently, the H.264 codec was selected for post-editing video rendering. The corpus of video files chosen for this experiment comprised ten videos, each with a duration ranging from a minimum of 30 seconds to a maximum of one minute.

\section{Evaluation}
\label{evaluation}

In this section, we present a comprehensive evaluation of our experiments. The evaluation is divided into two primary segments: Preliminary Data Exploration (Section \ref{subsec:preliminary-data-exploration}) and Classification Results (Section \ref{seubsec:classification-results-x}). This evaluation also presents reliable accuracy of the attack experiments and features impacting the experiments of CNN model architecture inference and the video inference used in video rendering tasks running on the GPU for both I/O ports (USB and HDMI).

\subsection{Preliminary Data exploration}
\label{subsec:preliminary-data-exploration}


In this study segment, we discuss how the raw data was prepared with feature extraction and removal of empty entries for classification. Initially, the data were captured in analog values form during the experiments. Afterward, each sample was stored as an array and subsequently organized into a dataframe. Missing or NaN values were removed after the data collection. In order to identify all CNN models from the collected power leakage data, we employed statistical and spectral feature extraction methods using Python's NumPy \cite{harris2020array} and Librosa \cite{mcfee2015librosa}, respectively. The extracted features, used in our experiments, both statistical and spectral, are outlined in Table \ref{table:features}.

\begin{table}[]
\centering
   \resizebox{0.49\textwidth}{!}{
\begin{tabular}{|c|c|}
\hline
\textbf{Feature Type} & \textbf{Extracted Features} \\ \hline \hline
Statistical Features  & \begin{tabular}[c]{@{}c@{}}Summation, Mean, Mean Absolute Deviation \\(MAD),Standard Deviation,  Variance, \\ Skewness, Standard Error of the \\Mean (SEM), Kurtosis\end{tabular}\\ \hline
Spectral Features     & \begin{tabular}[c]{@{}c@{}}Spectral Centroids (Multiple), Spectral\\ Bandwidths  (Multiple),  Spectral \\Flatnesses (Multiple), Spectral \\Roll-offs  (Multiple), Tempograms \\(Multiple), Spectral Contrasts (Multiple), \\Fourier Tempograms (Multiple), \\Zero Crossing Rate, Mel-frequency \\Cepstral Coefficients (Multiple),   \\Tempo, Chroma (Multiple), \\ Tempogram Ratio, Root Mean Square (RMS)\end{tabular} \\ \hline
\end{tabular}}
\caption{Statistical and Spectral features that were extracted from the collected data.}
\label{table:features}
\end{table}

\begin{figure*}[!htb]
     \centering
         \includegraphics[width=0.86\textwidth]{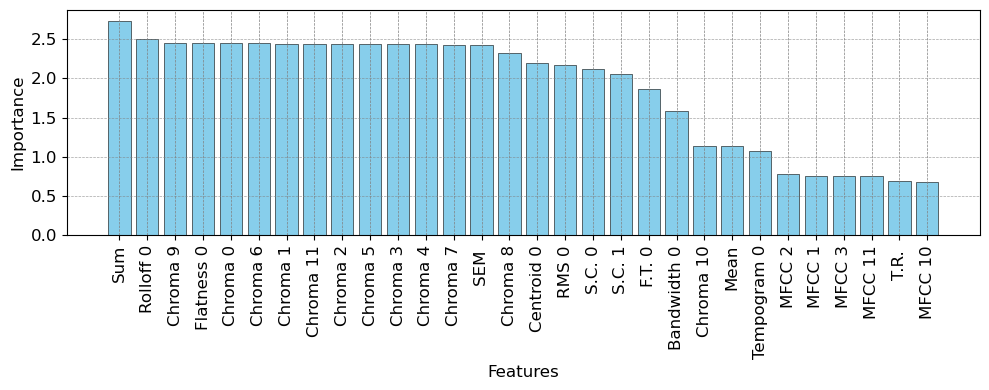}
        \caption{Feature importance from Scikit learn's Select K-Best for the top 30 features, derived from CNN models on Dataset-2. Abbreviations: F.T. (Fourier Tempogram), T.R. (Tempogram Ratio), S.C. (Spectral Contrast).}
        \label{feature-importance}
\end{figure*} 

\begin{figure*}[!ht]
    \centering
    \begin{subfigure}[b]{0.49\textwidth}
        \centering
        \includegraphics[width=\textwidth]{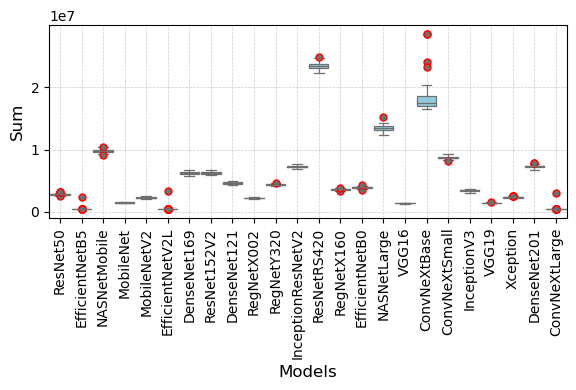}
        \caption{Feature: Sum}
    \end{subfigure}
    \hfill 
    \begin{subfigure}[b]{0.49\textwidth}
        \centering
        \includegraphics[width=\textwidth]{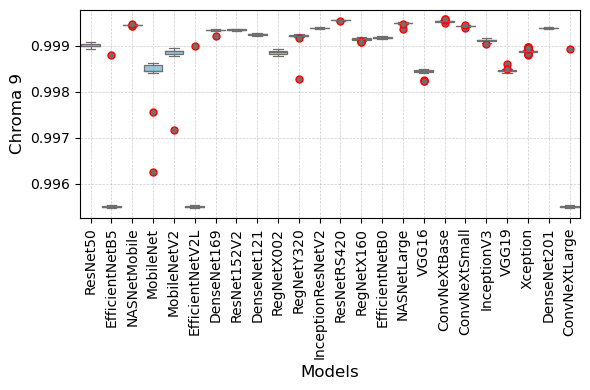}
        \caption{Feature: Chroma 9}
    \end{subfigure}
    \caption{Box plots presenting the top two features - Sum (Statistical feature) and Chroma 9 (Spectral feature) for all the CNN models involved in the experiments. The red dots represents the outliers.}
    \label{fig:features_some}
\end{figure*}

After the feature extraction process, we investigated the features that were extracted. For this part, we considered the features extracted using the Dataset-2 for visualization purposes. In order to identify the best features, we used the Scikit learn's \cite{pedregosa2011scikit} Select K-Best method for the top 30 features. First, we discuss the CNN models identification feature extraction process. We found that the feature sum and roll-off (1st roll-off) are of the highest importance in terms of both statistical and spectral features. It is important to mention that many spectral features produce more than one value for that specific based on the characteristics of the given time series. We also observed that chroma, flatness, SEM, and RMS are of higher importance. We presented the importance value created by the Select K-Best for the top 30 features in Figure \ref{feature-importance}. In addition, we investigated the values of some of the top features. In Figure \ref{fig:features_some}, we present the two main features - Sum (Statistical feature) and Chroma 9 (Spectral feature). This figure presents box plots of the values for all CNNs involved in the experiment for those two features. We can observe that some of the CNN models are distinguishable visually by observing these two features.

\begin{figure}[!ht]
    \centering
    \begin{subfigure}[b]{0.235\textwidth}
        \centering
        \includegraphics[width=\textwidth]{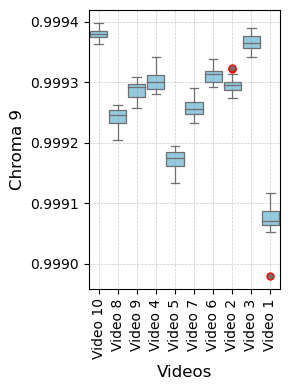}
        \caption{Chroma 9}
    \end{subfigure}
    \hfill
    \begin{subfigure}[b]{0.235\textwidth}
        \centering
        \includegraphics[width=\textwidth]{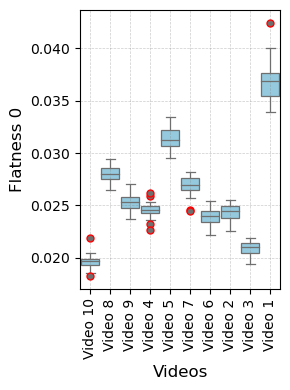}
        \caption{Flatness 0}
    \end{subfigure}
    \caption{Plots of Mean feature and MFCC 2 for Matrix Multiplication experiments.}
    \label{fig:features_video_edit}
\end{figure}

Afterwards, we discuss the feature extraction outcome of the video identification experiment. We found the top features to be Sum, Chroma 9, Flatness 0, Chroma 11 and Rolloff features. In this part, we visualize the values of two spectral features, Chroma 9 and Flatness 0. The values of these two features are presented in figure \ref{fig:features_video_edit}. It can be seen that Chroma 9 and Flatness 0 have visually distinguishable features for the 10 different videos.

\subsection{Classification}
\label{seubsec:classification-results-x}

In our research, we utilized the Scikit-learn library \cite{pedregosa2011scikit} to develop pipelines using various scalars and classifiers for data preprocessing and classification tasks, respectively. The scalars included: Max-Abs Scaler, ensuring the maximum absolute value of each feature is 1.0 without centering; Min-Max Scaler, adjusting features to the minimum and maximum range; Standard Scaler, normalizing features by removing the mean and scaling to unit variance; Quantile Transformer, mapping data to a uniform distribution and minimizing outlier effects; and Normalizer, scaling samples to unit norm, suitable for datasets with many zeros and varying scales. Classifiers employed were K-Neighbors \cite{10.1007/978-3-540-39964-3_62}, Light GBM \cite{10.5555/3294996.3295074}, Support Vector Machine (Multiclass) \cite{708428}, Random Forest \cite{Breiman2001}, and XGBoost \cite{10.1145/2939672.2939785}. We used the K-best algorithm to select the top 30 features for the pipelines, choosing the classifier with the highest test accuracy and the F1 score. We also set a confidence threshold of 0.5 for all classifiers to handle unknown class cases. Next, we discuss the classification results for the CNN and Video identification experiments in Section \ref{subsubsec:model-classification-results} and \ref{subsubsec:video-classification-results}.

\subsubsection{Classification results for CNN model architecture inference experiment}
\label{subsubsec:model-classification-results}
\begin{figure}[!htb]
     \centering
         \includegraphics[width=0.5\textwidth]{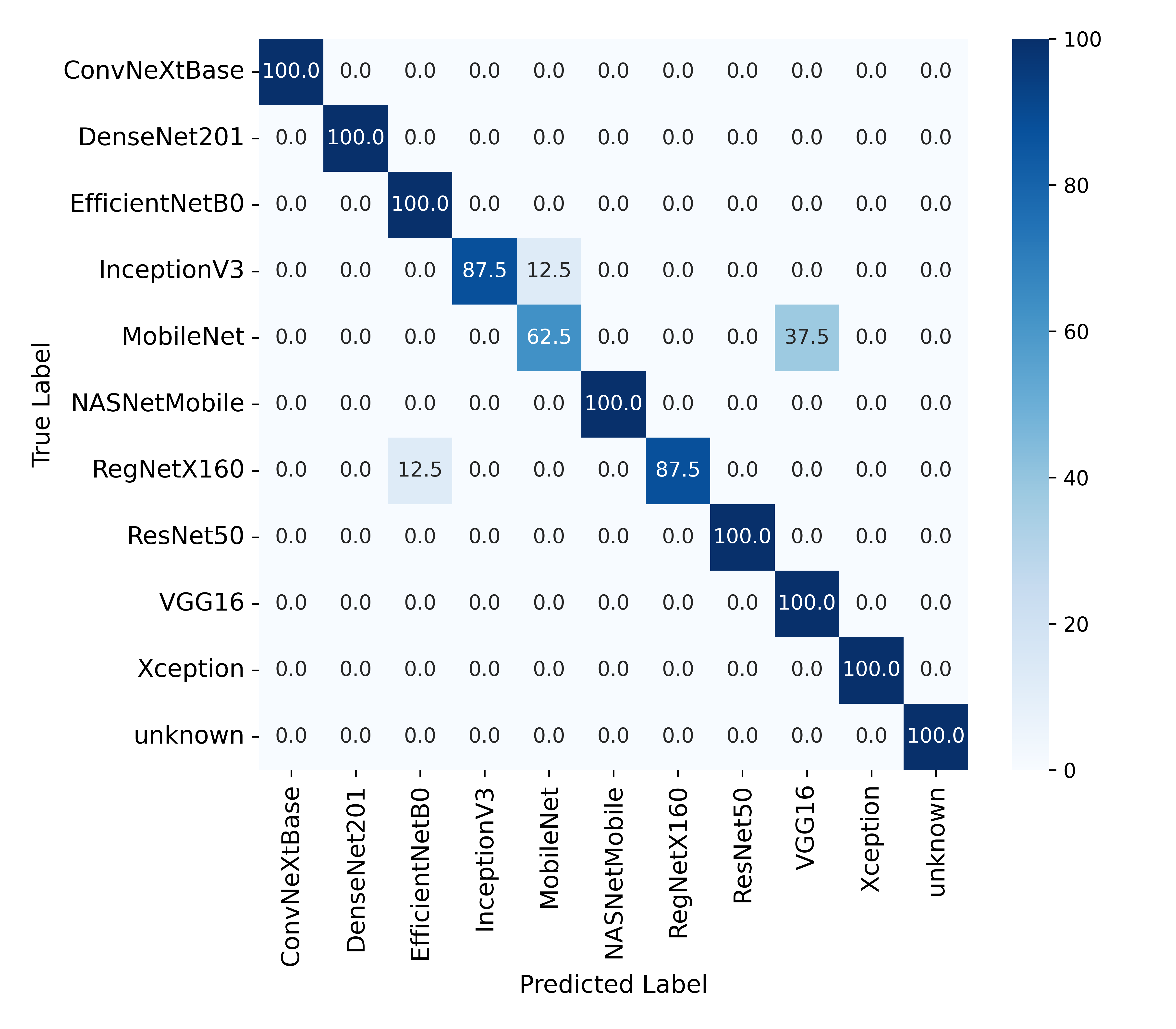}
        \caption{Confusion Matrix for the Classification of Base CNN models (Configuration 1), for the Dataset-2.}
        \label{cm-cnn-base-unknown}
\end{figure}

We first identify the base models of the networks mentioned in Table \ref{table:network_models} with multiclass classification. As mentioned before, we execute the sckit-learn pipeline composed of the scalar and the classifier. We run the classification step on both data sets (datasets 1 and 2) independently. This is formulated to be a 10-class problem. We conducted the classification task with the possibility of an unknown class and also without the possibility of having an unknown class. The test accuracy and the F1 score for the best performing model are given in Table \ref{tab:base-cnn}.
We can observe that, while monitoring the HDMI port, Dataset-2 had performed well with an accuracy of 94.25\% for the regular classification. Considering having an unknown class, the Dataset-2 performed better than the Dataset-1, having a test accuracy of 93.75\%. In both cases, the F1 score is higher. Furthermore, we present a confusion matrix for the Dataset-2 classification in Figure \ref{cm-cnn-base-unknown}. We can observe that in some of the test cases, MobileNet was misclassified as VGG16, InceptionNet was misclassified as MobileNet, and RegNet was misclassified as EfficientNet. Moreover, upon meticulous examination of the USB port, Dataset-2 exhibited a remarkable accuracy of 98.95\% when accounting for the potential presence of an unknown class. In contrast, when the scenario excluded the consideration of an unknown class, Dataset-1 demonstrated superior test accuracy, achieving an impressive 71. 35\%. 

Subsequently, we executed the classification for Configuration 2. It should be noted that, during the HDMI port monitoring, Dataset-2 exhibited commendable performance, achieving an accuracy of 76.45\% in standard classification scenarios. When incorporating an unknown class into the analysis (by setting the confidence threshold to 0.5 and providing unknown power leakage data), Dataset-2 outperformed Dataset-1, registering a test accuracy of 74.39\%. In both cases, the F1 scores were significantly higher, underscoring the robustness of the model. Furthermore, a thorough investigation of the USB port revealed that Dataset-2 maintained a remarkable accuracy of 74.39\% while considering the potential inclusion of an unknown class. In contrast, in scenarios where the unknown class was not considered, Dataset-2 showed superior test accuracy, attaining an impressive 72.10\%. The results are presented in Table \ref{tab:cnn-classificaiton-config-2}.


Across all instances examined, the K-Neighbors classifier emerged as the superior model, employing data normalization via the sklearn normalizer scaler. This model was meticulously configured through a pipeline designed to fine-tune hyperparameters. The optimized parameters included: the algorithm parameter was configured to 'auto' to enable automatic selection of the algorithm, leaf\_size was established at 30, the metric was designated as Minkowski with p=2 to denote Euclidean distance, and n\_neighbors was fixed at 5 with weights set to uniform. However, in some of the cases the XGBoost classifier performed well, such as Dataset-2, unknown class classification while monitoring the USB port. The random forest classifier performed better than the other classifiers when working with Dataset-1 and for regular classification (without considering the possibility of unknown class) while monitoring the USB port.

\begin{table}[]
\centering
\resizebox{0.5\textwidth}{!}{%
\begin{tabular}{|c|c|cc|cc|}
\hline
\multirow{2}{*}{\textbf{Classification}} & \multirow{2}{*}{\textbf{Dataset}} & \multicolumn{2}{c|}{\textbf{HDMI}} & \multicolumn{2}{c|}{\textbf{USB}} \\ \cline{3-6} 
                                        &           & \multicolumn{1}{c|}{Accuracy} & F1 Score & \multicolumn{1}{c|}{Accuracy} & F1 Score \\ \hline \hline
\multirow{2}{*}{\textbf{Regular}}       & Dataset-1 & \multicolumn{1}{c|}{87.36\%}  & 0.8747   & \multicolumn{1}{c|}{71.35\%}  & 0.7063   \\ \cline{2-6} 
                                        & Dataset-2 & \multicolumn{1}{c|}{94.25\%}  & 0.9415   & \multicolumn{1}{c|}{56.77\%}  & 0.5661   \\ \hline
\multirow{2}{*}{\textbf{Unknown}} & Dataset-1 & \multicolumn{1}{c|}{86.25\%}  & 0.8638   & \multicolumn{1}{c|}{98.34\%}    & 0.9814       \\ \cline{2-6} 
                                        & Dataset-2 & \multicolumn{1}{c|}{93.75\%}  & 0.9364   & \multicolumn{1}{c|}{98.85\%}  & 0.9885   \\ \hline
\end{tabular}%
}
\caption{Test accuracy for the Base CNNs classification (Configuration 1).}
\label{tab:base-cnn}
\end{table}

\begin{table}[]
\centering
\resizebox{0.5\textwidth}{!}{%
\begin{tabular}{|c|c|cc|cc|}
\hline
\multirow{2}{*}{\textbf{Classification}} & \multirow{2}{*}{\textbf{Dataset}} & \multicolumn{2}{c|}{\textbf{HDMI}} & \multicolumn{2}{c|}{\textbf{USB}} \\ \cline{3-6} 
                                        &           & \multicolumn{1}{c|}{Accuracy} & F1 Score & \multicolumn{1}{c|}{Accuracy} & F1 Score \\ \hline \hline
\multirow{2}{*}{\textbf{Regular}}       & Dataset-1 & \multicolumn{1}{c|}{68.34\%}   & 0.6829   & \multicolumn{1}{c|}{68.48\%}   & 0.6833   \\ \cline{2-6} 
                                        & Dataset-2 & \multicolumn{1}{c|}{76.45\%}   & 0.7667   & \multicolumn{1}{c|}{72.10\%}   & 0.7187   \\ \hline
\multirow{2}{*}{\textbf{Unknown}} & Dataset-1 & \multicolumn{1}{c|}{62.58\%}   & 0.6256   & \multicolumn{1}{c|}{60.50\%}   & 0.6014   \\ \cline{2-6} 
                                        & Dataset-2 & \multicolumn{1}{c|}{74.39\%}   & 0.7437   & \multicolumn{1}{c|}{71.39\%}   & 0.7321   \\ \hline
\end{tabular}%
}
\caption{Test accuracy for the CNN family classification (Configuration 2).}
\label{tab:cnn-classificaiton-config-2}
\end{table}


\subsubsection{Classification results for video rendering inference experiment}
\label{subsubsec:video-classification-results}
For the video identification task, we again perform a multiclass classification. We use the same machine learning pipelines discussed earlier in this classification task. In this experiment, the K neighbor classifier with the normalizer scalar also performed the best. 

\begin{figure}[!htb]
     \centering
         \includegraphics[width=0.5\textwidth]{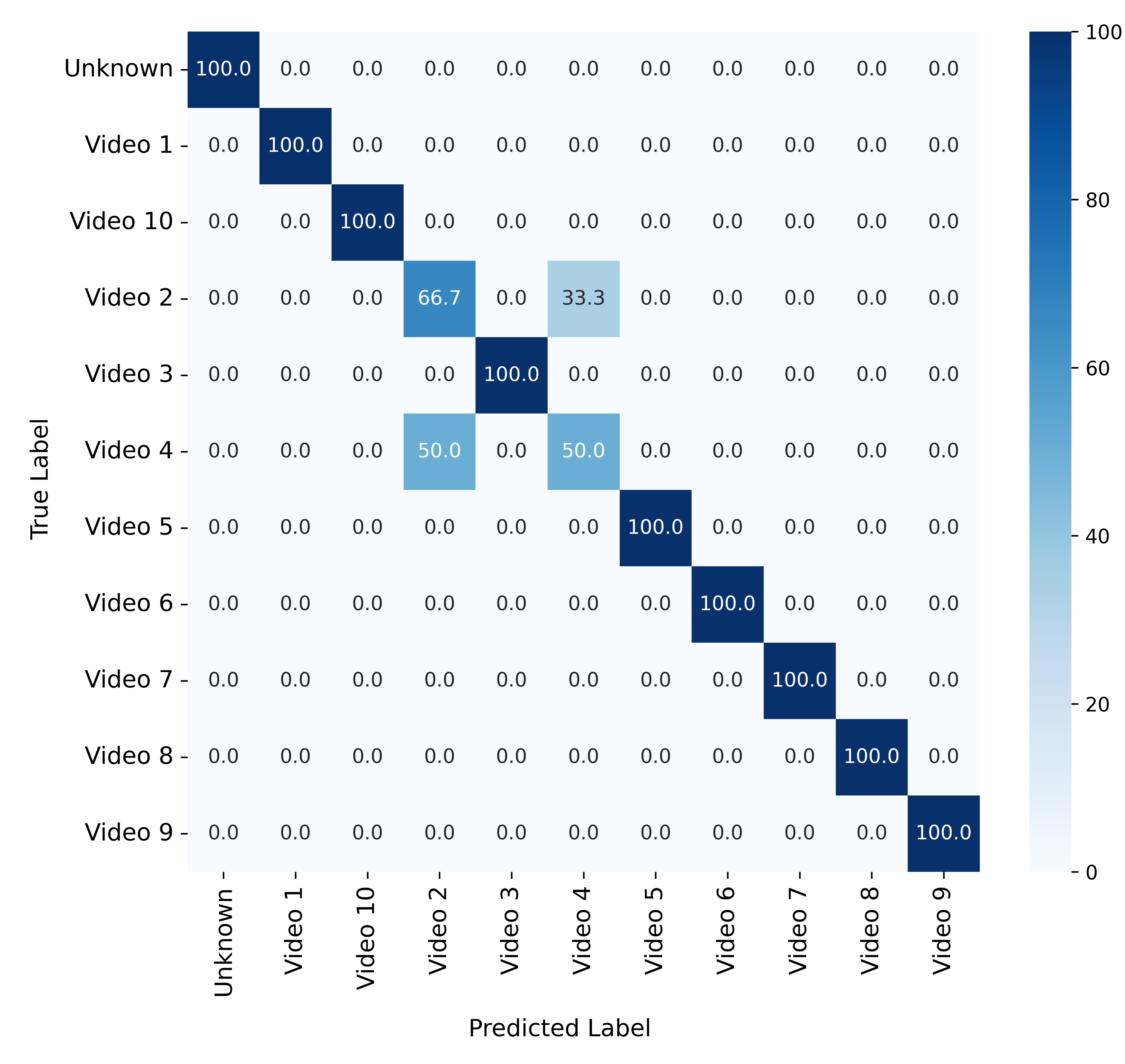}
        \caption{Confusion Matrix for the video rendering inference experiment.}
        \label{cm-video-editing-unknown}
\end{figure}

The test accuracy obtained in this classification task with the USB port is 92.86\%. The F1 score is 0.92. The scikit learn pipeline optimizes the classification task by setting the hyperparameters of the classifiers. In this experiment, the hyper parameters of the K neighbor classifier were set as follows. The leaf\_size is set to 30, metric as Minkowski with p = 2 representing the Euclidean distance, and n\_neighbors set to 5 with weights as uniform. We also present a confusion matrix for this experiment in Figure \ref{cm-video-editing-unknown}. We can observe that video 2 and video 4 were misclassified among themselves for a few test cases. On the other hand, with the identical classifier configuration, the test accuracy for the HDMI port achieved an impeccable 100\%. When we considered the possibility of having an unknown class by setting the confidence threshold to 0.5 and providing unknown power leakage data, the best performing average test accuracy was observed from Dataset-2 on both HDMI and USB ports, which were 74.39\% and 71.39\% respectively. The sklearn classification configuration remained same as before in this case.

\begin{table}[]
\centering
\resizebox{0.5\textwidth}{!}{%
\begin{tabular}{|c|cc|cc|}
\hline
\multirow{2}{*}{\textbf{Classification}} & \multicolumn{2}{c|}{\textbf{USB}} & \multicolumn{2}{c|}{\textbf{HDMI}}   \\ \cline{2-5} 
                 & \multicolumn{1}{c|}{Accuracy} & F1 Score  & \multicolumn{1}{c|}{Accuracy} & F1 Score\\ \hline \hline
\textbf{Regular} & \multicolumn{1}{c|}{92.86\%}      & 0.92 & \multicolumn{1}{c|}{100\%}     & 1  \\ \hline 
\textbf{Unknown}                        & \multicolumn{1}{c|}{96.00\%} & 0.9589 & \multicolumn{1}{c|}{96.43\%} & 0.9609 \\ \hline
\end{tabular}%
}
\caption{Classification results for the video identification experiment}
\label{tab:my-tablexxx}
\end{table}

\section{Limitation}
\label{limitation}
In this section, we delve into the critical limitations of the power side-channel, which can significantly influence the accuracy of the attack. We explore two pivotal limitations here: first, the necessity for the proximity of the attacker, where the attacker must be in close proximity to the victim (Section \ref{subsec:limitation-proximity}). Secondly, if the computer is executing multiple programs simultaneously, the attack may become infeasible (Section \ref{subsec:limitation-bnc}). Hence, the consideration of background noise is paramount.

\subsection{Attacker Proximity}
\label{subsec:limitation-proximity}

Our attack necessitates the attacker to meticulously read the current readings from the I/O port of the victim. This can be accomplished through either a USB cable directly linking the attack device or via wireless communication. Our sophisticated attack device can be seamlessly extended to utilize the WiFi or Bluetooth module, replacing the traditional wired USB connection between the victim and the attacker. Leveraging wireless communication empowers the attacker to significantly increase the distance from the victim, thereby minimizing the risk of detection. However, the maximum feasible distance is constrained by the coverage range of WiFi or Bluetooth technology. Notably, certain Bluetooth 5.3 version products boast an impressive effective range of up to 25 feet. In contrast, WiFi modules can span distances of up to 120 feet. Consequently, the attacker must be in close proximity to the victim to gather power side-channel data, necessitating a certain degree of nearness to the victim.

On another note, the attacker could modify the attack device to store the power side channel data from the victim onto a temporary storage medium, such as an SD card. This allows the attacker to later retrieve the SD card from the compromised device connected to the victim's machine. In such a scenario, the attacker must be physically present near the victim's machine not only to connect the attack device initially, but also to return and retrieve the collected data. This dual presence heightens the risk and audacity of the attack, showcasing the lengths to which an attacker might go to exploit vulnerabilities.

\subsection{Background Noise Consideration}
\label{subsec:limitation-bnc}
While conducting our experiments, we deliberately chose not to disable any system processes. We ensured that all operating system-related processes remained active. To maintain a controlled environment, we meticulously avoided any additional services or software from the user end, such as web browsers, games, etc. However, theoretically speaking, the presence of other running processes on the system could potentially interfere with or distort the attack. We rigorously tested several fundamental operations, including matrix multiplication, video rendering, and CNN execution. If a similar process were to run concurrently, it could indeed weaken the attack. Consequently, background noise can significantly impact the attack.
\section{Conclusion}
This paper unveils an innovative methodology for executing power side-channel attacks by redirecting the emphasis from the conventional practice of directly measuring the power consumption of processing units to scrutinizing peripheral components that share the same power supply. In particular, we explore the potential for harnessing power fluctuations in USB and HDMI ports as a novel vector for such attacks. This innovative mechanism can be used adeptly to extract valuable information from the I/O ports of aging computational devices. This opens new avenues for security research and exploitation.

We meticulously crafted a series of experiments to investigate the complicated operations of GPUs. Our initial focus was on unraveling the unique power leakage characteristics during simple matrix multiplication on a GPU. We found that by scrutinizing the power leakage, one can extract valuable insights about the matrix multiplication processes executed on the GPU. Building upon this revelation, we directed our research by orchestrating a comprehensive array of experiments. These experiments spanned diverse tasks, including the intricate realms of video rendering and the sophisticated architectures of convolutional neural networks (CNNs), to meticulously infer the underlying architecture of CNN models.

We structured our experiments to identify the family of CNN models that are used on the GPU and further tried to identify the variant of the CNN model that is used on the GPU. Additionally, we explored the potential for identifying hitherto unknown classes within our experimental framework. In our experiments, we investigated two widely used I/O ports, such as USB and HDMI ports in all our experiments.

Our findings demonstrate that power fluctuations in peripheral components can indeed be harnessed to reveal significant information about GPU operations such as video rendering, matrix multiplication, and the architecture of neural network models. This not only broadens the scope of power side-channel attacks but also underscores the need for comprehensive security measures that encompass all aspects of the power supply network. By shifting the focus from direct power measurement of processing units to peripheral components, we provide a new perspective on the potential vectors for power side-channel attacks.

The power leakages enabling our attack could be due to several reasons. First of all, computer power supply units (PSU) stabilize the electrical current of system components using capacitors, MOSFETs and voltage regulation modules (VRMs). These components ensure consistent current flow, protecting against performance or power fluctuations \cite{vrms_power}. However, due to wear and tear the effectiveness of power supply stabilization may deteriorate over time. This could contribute to the power draw patterns that our attack picks up at the USB and HDMI ports. Another possible reason could be that the power stabilization mechanisms of today's computers are simply not fine-grained enough to obfuscate the minute power fluctuation patterns that our algorithms leverage to drive the attacks. A thorough exploration on the potential contributions of these variables is part of our future work. 

The implications of this study offer new insights and opportunities for both attackers and defenders in the ongoing battle to secure computational devices.

\bibliographystyle{IEEEtran}
\bibliography{references}

 
\newpage

 




\vfill

\end{document}